# 사이버테러를 동반한 테러 양상 예측 및 관련 법제도 발전방향

김대건

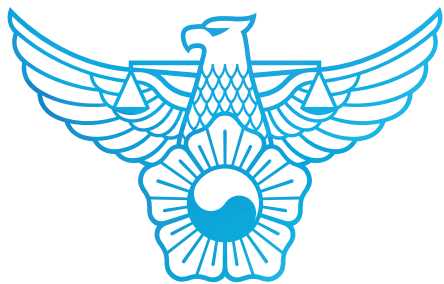

# 사이버테러를 동반한 테러 양상 예측 및 관련 법제도 발전방향
Prediction of terrorism pattern accompanied by cyber-terrorism and the development direction of corresponding legal systems

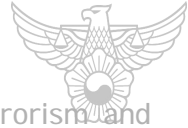

김 대 건*

차 례

Ⅰ. 서 론  
Ⅱ. 테러 양상 변화 예측  
Ⅲ. 사이버테러 대응을 위한 법제도 실태 및 문제점  
Ⅳ. 발전방향  
Ⅴ. 결 론

## 국문요약

국가 운영 및 국민 생활에 정보통신체계가 필수 요소로 자리 잡고, 국가기반체계 및 시설 등의 정보통신체계 의존성이 높아지면서 사이버테러는 평시 국가안보에 대한 심대한 위협요소로 급부상하고 있다. 테러집단의 사이버공격 자산에 대한 접근성이 향상되면서 전통적인 테러형태 또한 사이버테러와 결합된 형태로 변화될 것으로 예상된다. 그럼에도 불구하고, 우리나라는 국가안보적 관점에서 사이버테러에 대비하고 대응하기 위한 법제도가 미비한 실정으로, 사이버테러라고 일반적으로 지칭되는 과거에 발생한 사이버공격을 국가적으로 테러로 규정하고 대응하는데도 어려움이 있다. 본 논문에서는 현대 군사작전개념의 발전과정을 토대로 앞으로 발생 가능한 테러 양상의 변화를 예측하고, 현행 법제도를 기반으로 사이버테러에 대한 국가적 대응의 제한사항을 분석한 후. 향후 발전방향을 모색한다.

➲ 주제어  
사이버테러, 테러 양상 변화, 국가안보, 법제도, 테러방지법, 사이버안보

* 군사안보지원사령부 팀장, 고려대학교 정보보호대학원 공학박사  
  본 논문은 저자 개인의 견해이며 소속기관의 입장을 반영한 것이 아님.

# Ⅰ. 서 론

UN의 발표에 의하면 약 190개 가입 국가 중 우리나라의 전자정부 발전 순위는 2020년 기준으로 2위를 기록하였으며, 2010년 이후로 지속 3위 이내의 순위를 유지하고 있다.[1] 또한, 국제전기통신연합(International Telecommunication Union, ITU)에서 2017년에 발표한 국가별 ICT 발전지수에서는 우리나라가 2위를 기록하였다.[2] 이처럼 우리나라의 정보통신기술 발전은 세계적 수준이며 국가운영 및 국민 생활에 필수적인 요소로 자리 잡고 있다. 그러한 만큼, 정보통신체계가 서로 연결되어 다양한 정보를 저장, 유통 및 처리하는 사이버공간에 대한 의도된 위협은 파급효과가 상당할 뿐 아니라, '국민보호와 공공안전을 위한 테러방지법' (이하 '테러방지법'이라고 한다)에 정의된 '국가·지방자치단체의 권한행사를 방해하거나 의무 없는 일을 하게 할 목적 또는 공중을 협박할 목적'으로 하는 '테러' 행위에 해당하는 목적을 달성하기 위한 수단으로 이용되거나 테러 대상이 될 수 있다.

〈표 1〉의 경찰청의 공식 발표[3]에서 확인할 수 있듯이 사이버 범죄도 지속 증가하고 있는 추세이고, 많은 국가들은 이미 사이버 공격 수단을 국가의 전략적 목표를 달성하기 위한 수단으로 광범위하게 활용하고 있다. 이는 사이버 공간이 단순히 범죄 행위가 발생하는 공간으로 폭넓게 활용될 뿐 아니라, 다양한 사이버 공격 수단이 개발되고 유통됨으로써 범죄 집단이 사이버 공간에 접근하기 용이해진 것이 주된 이유이다.

〈표 1〉 사이버범죄 비율 및 증감 추이(2016-2020)

| 구 분 | | 2016 | 2017 | 2018 | 2019 | 2020 |
|---|---|---|---|---|---|---|
| 사이버 범죄 | 건 수 | 153,075 | 131,734 | 149,604 | 180,499 | 234,098 |
| | 증감률 | - | -13.94 | 13.57 | 20.65 | 29.69 |
| 전체 범죄 | 건 수 | 1,849,460 | 1,662,341 | 1,580,751 | 1,611,906 | 1,588,400 |
| | 증감률 | - | -10.12 | -4.91 | 1.97 | -1.46 |
| 사이버범죄 비율 | | 8.28 | 7.92 | 9.46 | 11.20 | 14.74 |

---

1) e-나라지표 'UN 전자정부평가'에서 발췌, https://www.index.go.kr/potal/main/EachDtlPageDetail.do?idx_cd=1027
2) 2017년 이후에는 순위가 발표되지 않았으며 세부 내용은 아래 링크 참조.
   (https://www.itu.int/net4/ITU-D/idi/2017/index.html#idi2017rank-ta)
3) 경찰청 국가수사본부 사이버수사국, "2020 사이버범죄 동향 분석 보고서", p.5

이렇듯, 사이버테러 (사이버안보)는 평시 국가안보에 대한 위협을 초래할 수 있는 매우 가능성 높은 행위임에도 불구하고, 기존의 연구들은 사이버안보를 정보통신체계 보호 수준으로 인식하는데 그쳐, 국가안보 보장을 위해 사이버테러에 대한 적극적 대응 활동을 검토하는데 한계를 보이고 있으며, 사이버테러를 사이버 범죄와 동일한 수준의 개념으로 인식한 상태에서 관련 법제도를 분석하고 있다. 본 논문에서는 국가안보의 일환으로서 사이버안보를 평시 단계에서 보장하기 위한 현 법제도의 문제점을 살펴보고 발전방향을 제시한다.

이에 앞서, 다음 장에서는 향후 테러 양상의 변화 방향을 예측함으로서 사이버테러 대비의 중요성을 강조하고자 한다.

## Ⅱ. 테러 양상 변화 예측

테러는 평시 국가안보를 위협하는 주된 위협요소이다. 따라서 테러 양상의 변화를 예측함으로써 이에 대한 대응역량을 발전시기 위해 집중해야 할 방향을 살펴볼 수 있다. 테러는 테러 행위자가 물리적 수단을 활용하여 테러 대상에 위력을 가한다는 면에서 군사작전과 유사성이 있다. 본 장에서는 군사작전 수단의 변화에 따라 새롭게 등장하는 작전개념의 변화 과정을 통해 향후 테러 양상의 변화를 예측한다.

### 1. 현대 군사작전개념 발전

전쟁사에 등장하는 모든 군사작전개념을 설명하는 것은 본 논문의 목적에서 벗어난다. 따라서, 본 절에서는 군사작전 수단이 변화함에 따라 작전개념도 함께 변한다는 것을 이해 할 수 있는 수준에서 설명하고자 한다. 또한, 사이버 공간 및 수단이 군사적으로 활용되면서 등장한 군사작전개념을 함께 설명한다.

'91년에 발발한 걸프전은 첨단 무기의 위력을 보여준 현대전의 시초라고 여겨진다. 미국이 주도하는 다국적군과 이라크 간의 전쟁인 걸프전에서는 F14 (Tomcat), F15 (Eagle), F114A (Nighthawk), B52G (Stratofortress) 등 당대 최신 전투기 및 폭격기, AH-64 (Aphache) 등 회전익 항공기와 순항미사일 등이 동원되어 다국적군이 재래식 전력 위주인 이라크에게 압도적인 승리를 거두었다. 과거 전장에서의 전선 (前線)은 지상을 중심으로 한 선형 (線型)의 개념으로 여겨져 왔다. 그러나, 위와 같은 무기체계의 발전으로 인해 전장은 더 이상 선형적인 전선으로

국한되지 않고 적 후방의 종심지역으로 전선을 동시에 확대하여 비선형적인 공간에서 전투를 수행할 수 있게 됨으로써 공지전투 (Airland battle)의 작전개념을 구현할 수 있게 되었다.

걸프전에서도 네트워크 중심의 지휘통제체계는 전쟁의 승리를 보장하기 위한 핵심적인 요소였으나, 지원수단으로서의 성격이 더욱 강했다. 그러나, 점차 네트워크 공간의 개념이 사이버 공간으로 확장되고, 사이버 공간이 지상, 해상, 공중, 우주공간과 함께 제 5의 전장으로 인식되면서 독자적인 작전 공간으로 활용되게 되었다. 러시아-조지아 분쟁 (남오세티아전 또는 그루지야전, 2008)은 사이버전 (Cyber warfare) 및 하이브리드전 (Hybrid warfare)으로 불리는 대표적인 사례이다. 러시아와 조지아의 본격적인 물리적 충돌에 앞서 조지아의 정부 홈페이지, 언론사 등이 대규모 분산 서비스 거부 (DDoS) 공격이 받았으며, 이는 물리전과 사이버전이 병행될 수 있음을 보여준 대표적인 사례로 꼽히고 있다.4)

이후 사이버 공간에서의 군사력은 공통의 군사작전 (전략) 목표 달성을 위해 지상, 해상, 공중, 우주 영역에서의 군사력과 시간적으로 통합되거나, 특정한 전략적 상황에 도달하기 위한 과정에서 다른 공간에서의 군사력 운용과 유기적으로 통합되는 방향으로 발전하였으며, 미군에서는 이를 다영역작전 (Multi-domain operations)으로 정립하였다.

다영역작전은 2018년 미국의 국방 전략이 전 세계 폭력 극단주의자들에 대항하는 것에서 러시아 및 중국 중심의 수정주의 세력에 대항하는 방향으로 수정됨에 따라 이를 달성하기 위해 미 육군을 중심으로 발전시킨 군사작전개념이다. 미군은 중국의 대미 군사전략인 반접근/지역거부 (Anti-Access/Area Denial, A2/AD)와 같은 적국의 군사전략에 대항하여 전략목표를 달성하기 위해 군사작전 단계를 '경쟁 (Competition) - 침투 (Penetration) - 분리 (Disintegration) - 확대 (Exploitation) - 재경쟁 (Recompetition)'으로 구분하였다.5) 전통적으로 군사작전이 수행되는 공간은 지상, 해상 및 공중으로 여겨져 왔으나, 다영역작전에서는 〈그림1〉6)과 같이 각 작전단계를 성공으로 이끌기 위한 결정적 지점에서 우주 및 사이버 공간까지 합동작전 영역으로 통합하게 된 것이다.

---

4) 육군사관학교 교수 등 공저, 사이버전 개론, 양서각, 2013, p.83
5) 조상근, "다영역작전 (Multi-Domain Operations, MDO) – 지상, 해상, 공중, 우주, 사이버·전자기 영역을 종횡무진 (縱橫無盡)하는 미 육군의 미래 작전수행개념",
　https://bemil.chosun.com/site/data/html_dir/2020/12/31/2020123100581.html (검색일: 2021.10.24.)
6) U.S. Army Training and Doctrine Command, Pamphlet 525-3-1: The U.S. Army in Multi-Domain Operations 2028, 2018, p.26

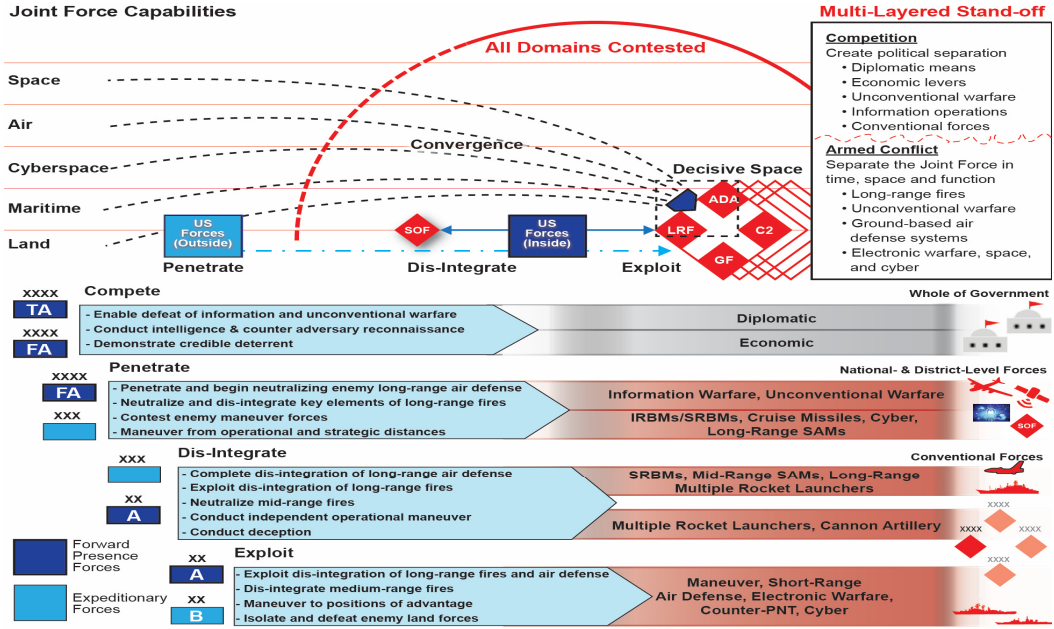

〈그림 1〉 다영역작전 수행 개념 및 사이버공간 (수단) 활용

　〈표1〉에서 수치로 확인한 것과 같이 사이버 공간은 이미 범죄의 수단 및 대상으로 활발하게 이용되고 있으며, 사이버 범죄 유형 중 수행 방식 및 영향력 측면에서 사이버 테러와 가장 유사한 '침해범죄'[7]는 '20년 기준으로 전체 유형 중 1.9% (4,344건)에 불과하지만 '19년 대비 증가율은 19.4%을 보여 타 유형의 증가율과 유사한 수준을 나타내고 있다. 이는 과거 사이버 침해 (공격) 수단 및 관련 정보는 전문가의 전유물이었으나 다크 웹 (Dark Web), 딥 웹 (Deep Web) 등을 통한 침해 (공격) 수단의 공유 및 판매 등이 활발해지면서 이에 접근할 수 있는 문턱이 점점 낮아진 결과로 볼 수 있다. 일례로, 2000년 초반부터 랜섬웨어 (Ransomware)가 사이버 범죄 수단으로 부상하면서 수많은 유형의 랜섬웨어가 등장하였고, 단순한 금전적 피해에서부터, 정보 유출, 인명피해 및 국가적 피해 등 크고 작은 피해가 발생하였다.[8] 랜섬웨어에 의한 피해가 지속되는 배경으로 랜섬웨어 서비스 (Ransomware as as Service, RaaS)의 등장을 들 수 있다. 랜섬웨어 제작을 위한 전문적인 지식 없이도 이를 구매하여 유포한 후 얻은 수익을 제작자와

---

7) 경찰청에서는 사이버범죄를 정보통신망 침해범죄(해킹, 악성프로그램 유포 등 정보통신망에 불법적으로 침입하는 방식으로 저지르는 범죄), 정보통신망 이용범죄(사이버사기, 사이버금융범죄 등 정보통신망을 이용해 저지른 범죄), 불법콘텐츠 범죄(사이버성폭력, 사이버도박 등 법이 금지하는 정보 등을 생산·유포하는 범죄)로 구분하고 있음.
8) 민경식, 김영직, 박진상, 장한나, 랜섬웨어 최신동향 분석 및 시사점, KISA Insight, 한국인터넷진흥원, vol. 2, 2021, p.9

공유하는 방식, 랜섬웨어의 유포 서비스까지 제공하는 방식, 방어 시스템에 의한 탐지가 어렵도록 랜섬웨어의 업데이트까지 제공하는 방식 등9) 다양한 유형으로 서비스가 제공되어 공격 수단에 대한 접근성이 폭증하게 된 것이다. 이렇듯, 일반인의 사이버 범죄 수단에 대한 접근성은 점점 증대되고 있으며, 향후에는 단순히 취약점 실행 코드나 유출된 정보를 판매하는 것이 아닌 네트워크 접근 권한을 판매하여 공격의 문턱을 낮아지고 공격 규모가 확대될 것으로 예상된다.10)

이러한 경향은 테러집단에게도 사이버 공격 수단에 대한 접근성을 확대시키고 접근 비용도 점차 낮아져, 테러 유형을 전통적 테러 수단과 사이버 수단이 결합된 군사작전개념의 변화와 유사한 형태로 발전시킬 것이다. 즉, 테러집단들은 앞으로 항공, 선박, 화생방, 폭발물 등 물리적 수단을 주로 활용했던 전통적인 테러 형태에서, 다영역작전과 같이 테러 목적을 달성하기 위해 전통적 수단과 사이버 공격 수단을 결합시켜, 전통적 수단에 의한 목적 달성 가능성을 향상시키거나, 테러의 파급 수준, 범위 및 지속성 등을 폭증시키고자 할 것이다.

## Ⅲ. 사이버테러 대응을 위한 법제도 실태 및 문제점

우리나라에서 사이버테러를 법률로 정의하려는 시도는 있었으나, 관련 법률이 제정된 사례는 아직 없다. 따라서, 본 장에서는 사이버 공간에서의 위협 행위를 사이버테러, 사이버공격, 사이버범죄 등으로 구분하지 않고 사이버침해와 관련된 법제도 실태를 포괄적으로 알아볼 것이며, 법률상으로 명시된 '테러'의 목적과 동일한 목적11)으로 수행된 사이버침해를 '사이버테러'로 잠정하여 이에 대한 대응상의 문제점을 살펴보려 한다.

### 1. 사이버테러 관련 법제도

본 절에서는 사이버테러 관련 법제도의 목적, 적용대상 및 범위, 실제 침해 행위 발생 시 수행 가능한 대응활동의 범위 및 주체를 중심으로 살펴본다.

---

9) 한수연, [IT열쇳말] 서비스형 랜섬웨어(RaaS), 블로터,
   https://www.bloter.net/newsView/blt201708170001 (검색일: 2021.11.7.)
10) 한국인터넷진흥원, 2021년 사이버 위협 전망, 2021, p.6,
   https://www.krcert.or.kr/data/reportView.do?bulletin_writing_sequence=35878 (검색일: 2021.11.7.)
11) 테러방지법 제2조(정의) 1항에서는 '테러'를 국가・지방자치단체 또는 외국정부의 권한행사를 방해하거나 의무 없는 일을 하게 할 목적 또는 공중을 협박할 목적으로 하는 특정 행위로 정의하고 있다.

### 가. 정보통신기반 보호법

본 법률은 '전자적 침해행위에 대비하여 주요정보통신기반시설[12]의 보호에 관한 대책을 수립·시행함으로써 동 시설을 안정적으로 운용하도록 하여 국가의 안전과 국민생활의 안정을 보장'하기 위한 것으로 중앙행정기관별 소관분야의 정보통신기반시설 중에서 주요정보통신기반시설을 지정하여 보호하도록 명시하고 있으며, 주요정보통신기반시설에 대한 광범위한 침해 발생 시 국무조정실장을 위원장으로 하는 정보통신기반보호위원회에 정보통신기반 침해사고 대책본부를 구성하여 응급대책, 기술지원 및 피해복구 등을 수행할 수 있다고 규정하고 있다.

### 나. 정보통신망 이용촉진 및 정보보호 등에 관한 법률 (이하 정보통신망법)

본 법률은 '정보통신망[13]의 이용을 촉진하고 정보통신서비스[14]를 이용하는 자를 보호함과 아울러 정보통신망을 건전하고 안전하게 이용할 수 있는 환경을 조성'하기 위한 것으로서, 정보보호 측면에서는 정보통신서비스 제공자의 정보통신망 안정성 확보를 위한 의무사항을 중심으로 규정하고 있다. 또한, 침해사고 발생 시 신고의무를 부여하고 있으며, 과학기술정보통신부장관은 중앙행정기관의 장과 협력하여 원인분석, 피해확산 및 재발방지 등의 대응행위를 할 수 있다고 규정하고 있다.

### 다. 국가사이버안전관리규정

이 규정은 행정규칙 (대통령훈령)으로, '국가사이버안전에 관한 조직체계 및 운영에 대한 사항을 규정하고 사이버안전업무를 수행하는 기관간 협력을 강화함으로써 국가안보를 위협하는 사이버 공격으로부터 국가정보통신망을 보호'하기 위해 제정되었다. 규정의 적용범위는 '중앙행정기관, 지방자치단체 및 공공기관의 정보통신망'이나, 정보통신기반보호법에 따른 주요정보통신기반시설에 대해서는 정보통신기반보호법을 우선 적용하도록 명시하고 있다.

---

[12] "정보통신기반시설"이라 함은 국가안전보장·행정·국방·치안·금융·통신·운송·에너지 등의 업무와 관련된 전자적 제어·관리시스템 및 「정보통신망 이용촉진 및 정보보호 등에 관한 법률」 제2조 제1항 제1호에 따른 정보통신망을 의미한다.
[13] "정보통신망"이란 「전기통신사업법」 제2조 제2호에 따른 전기통신설비를 이용하거나 전기통신설비와 컴퓨터 및 컴퓨터의 이용기술을 활용하여 정보를 수집·가공·저장·검색·송신 또는 수신하는 정보통신체제를 말한다.
[14] "정보통신서비스"란 「전기통신사업법」 제2조 제6호에 따른 전기통신역무와 이를 이용하여 정보를 제공하거나 정보의 제공을 매개하는 것을 말한다.

이해를 돕기 위해 쉽게 설명하자면, 본 규정의 적용범위에 해당하는 정보통신망에 대하여 행위의 주체가 중앙행정기관장 및 국가정보원장으로 상이해진 것 이외에는, 정보통신망법에 정보통신망의 안전성 확보를 위해 규정된 정보통신서비스 제공자 및 과학기술정보통신부장관의 의무 및 권한 등과 유사한 내용을 규정하고 있으며, 각 중앙행정기관, 지방자치단체 및 공공기관의 정보통신망에 대하여 정보통신기반 보호법에 규정된 것과 유사한 보호대책을 수립 및 시행할 것을 명시하고 있다.

### 라. 사이버안보 업무규정

본 규정 (대통령령)은 국정원법에 명시된 국정원의 직무 중 '사이버안보 관련 정보의 수집, 작성, 배포 및 사이버공격, 위협에 대한 예방 대응 업무의 수행에 필요한 사항을 규정'하는 것으로, 중앙행정기관 등을 대상으로 하는 사이버 공격·위협의 탐지·대응을 위한 위협정보수집 및 공유, 보안관제, 경보발령, 공격 주체 규명, 원인 분석 및 피해 내역 확인 등을 위한 사고조사 등의 대응행위를 명시하고 있다.

### 마. 통합방위법

본 법령은 '적의 침투, 도발이나 그 위협에 대응하기 위하여 국가 총력전의 개념을 바탕으로 국가방위요소를 통합, 운용하기 위한 통합방위 대책을 수립, 시행하기 위하여 필요한 사항을 규정'하는 것을 목적으로 한다. 주로 전통적인 물리적 공간에서의 '적'의 침투 및 도발 등에 대한 대응을 위해 제정된 법으로써, 특별시·광역시·특별자치시·도·특별자치도 등의 물리적 (지리적) 공간을 기준으로 통합방위태세가 발령되고, 통합방위작전의 관할 구역을 지상·해상·공중 관할구역으로 구분하는 것으로 규정하고 있다.

### 바. 테러방지법

본 법령은 '테러의 예방 및 대응 활동 등에 관하여 필요한 사항과 테러로 인한 피해보전 등을 규정함으로써 테러로부터 국민의 생명과 재산을 보호하고 국가 및 공공의 안전을 확보하는 것을 목적'으로 하고 있으며. 테러 행위의 대상 및 수단을 '사람, 항공기, 선박, 생화학·폭발성·소이성(燒夷性) 무기나 장치, 핵물질' 등으로 규정하고 있다. 또한, 대테러활동을 '정보의 수집,

테러위험인물의 관리, 테러에 이용될 수 있는 위험물질 등 테러수단의 안전관리, 인원·시설·장비의 보호, 국제행사의 안전확보, 테러위협에의 대응 및 무력진압 등 테러 예방과 대응에 관한 제반 활동'으로 포괄적으로 명시하면서, 이에 필요한 전담조직을 설치할 수 있도록 규정하고 있다.

### 사. 사이버안보 관련 법안 발의

그동안 〈표 2〉와 같이 사이버안보와 관련된 다양한 법안이 발의되어왔으나 아직 입법된 사례는 없다. 특히, 이러한 유사 법안의 입법이 되지 않고 있는 이유를 2016년 입법 발의된 국가 사이버테러 방지 등에 관한 법률안에 대한 입법 타당성 평가연구에서 제시한 주요 쟁점15)을 통해 살펴보면, 기존 법률과 중복규제로 인한 비효율 발생가능성, '사이버안보법안'이라는 법안 명칭이 주는 국가 안보적 영향성에 비해 법안에서는 사이버보안 위협을 지나치게 협소하게 정의, 사이버보안(테러) 등 주요 개념의 지나친 포괄성, 국가정보원의 권한확대에 대한 우려 등이 있다. 사이버안보 관련 법안의 필요성에 대한 찬반 의견이 공존하는 가운데 관련된 논의는 끊이지 않고 최근에도 진행되고 있다.16)

## 2. 문제점

현행 법제도 속에서는 사이버테러에 대한 제한적인 대응만 가능하다. 앞서 살펴본 바와 같이 현행 법제도는 주로 사이버 침해 행위에 대한 법적 책임 및 위반시 제재 등 사법적으로 규정하고 있어, 사이버 범죄행위를 규정하고 이를 처벌 및 대응하기 위한 법규가 주를 이룬다. 물론, 이에 대한 대비 및 대응을 위한 관련 기관의 의무 및 책임, 대응 범위를 규정한 법률도 존재하나, 규정된 대응 범위가 사고조사(원인규명) 및 복구 수준에 그치고 있다.

테러방지법에는 대테러 행위의 범주에 무력제압이 포함되어있으나, 현존하는 사이버 테러 관련 법안에는 공격 행위자 및 자산에 대한 비례적 대응 권한 및 책임을 규정하지는 않아 적극적 대응이 제한되고, 의무 및 책임소재가 불명확하다.17) 또한, 테러방지법에는 사이버테러를 테러의 유형으로 명문화하여 포함되어있지 않아 사이버테러에 대한 대응을 위해 테러방지법을 적용하기는

---

15) (재)더미래연구소, 사이버테러방지법의 주요 쟁점분석을 통한 입법 타당성 평가, 국회정보위원회 정책연구개발용역 성과물, 2016, pp.16-20.
16) 박대로, "여야, 국정원에 광범위 사이버정보 수집권 부여 입법", 뉴시스 (2021.11.19.)
17) 사이버 공간의 연결성으로 인해 대응 역량이 국외로 투사될 경우의 탈린 메뉴얼과 같은 국제법적 고려사항은 논외로 한다.

제한된다.

〈표 2〉 사이버안보 관련 발의 법안

| 법 안 | 대표발의자 | 발의년도 |
| --- | --- | --- |
| 사이버위기 예방 및 대응에 관한 법률안 | 공성진 의원 | 2006 |
| 국가 사이버위기관리법안 | 공성진 의원 | 2008 |
| 국가 사이버안전 관리에 관한 법률안 | 하태경 의원 | 2013 |
| 국가 사이버테러 방지에 관한 법률안 | 서상기 의원 | 2013 |
| 사이버위협정보 공유에 관한 법률안 | 이철우 의원 | 2015 |
| 사이버테러 방지 및 대응에 관한 법률안 | 이노근 의원 | 2015 |
| 국가 사이버테러 방지 등에 관한 법률안 | 서상기 의원 | 2016 |
| 국가 사이버안보에 관한 법률안 | 이철우 의원 | 2016 |
| 국가사이버안보법안 | 정 부 | 2017 |
| 사이버안보 기본법안 | 조태용 의원 | 2020 |
| 국가사이버안보법안 | 김병기 의원 | 2021 |

한편, 통합방위법은 '적'에 의한 침투가 명확할 경우 적용할 수 있으나 사이버 공격 특성상 '적'을 규명하는데 시간이 소요되고, 명확히 규정하기 어려운 경우도 있으며, 정치적 상황에 따라 '적'에 의한 공격으로 판단됨에도 불구하고 이를 공표하지 못하는 경우도 발생할 수 있다. 또한, 통합방위법은 전통적인 물리공간에서 국토에 대한 적의 침해에 대응하기 위한 법으로, 침해된 물리적 공간을 구획하여 특정 태세를 발령 및 제 능력을 통합하여 대응하도록 개념이 규정되어있으며, 이로 인해 물리적 경계가 모호한 사이버공간에서 발생하는 사이버테러에는 적용이 제한된다.

무엇보다 어떠한 법안에도 '사이버테러'의 개념이 정의되어있지 않아, 사이버침해와 혼돈되거나, 이와 유사한 수준으로 인식되고 있다는 것이 큰 문제이다. 사이버테러로 인한 사이버공간에서의 기밀성, 무결성, 가용성 측면의 피해는 오프라인 공간에서 발생하는 테러행위로 인한 피해와 같이 가시적이지 않을 수 있으나, 잠재적인 영향성 및 피해범위 측면에서 국가안보 차원의 막대한 손실을 초래할 수 있다. 그럼에도 불구하고, 현재는 국가사이버안전규정에 '사이버위기'를

'사이버공격으로 정보통신망을 통해 유통·저장되는 정보를 유출·변경·파괴함으로써 국가안보에 영향을 미치거나 사회·경제적 혼란을 발생시키거나 국가 정보통신시스템의 핵심기능이 훼손·정지되는 등 무력화되는 상황'으로 사이버테러와 유사한 맥락으로 정의되어있는 수준이다.

## Ⅳ. 발전방향

기존의 다수의 연구에서는 사이버테러 역량과 관련 법적 기반을 강화하기 위한 <표 2>와 같은 사이버테러 (안보) 관련 독립적 법안의 필요성을 주장하고 있다. 그러나, 앞서 언급한 바와 같이 유사한 법률안이 오랜 기간 발의되어 왔음에도 나름의 논리로 인해 입법되지 않고 있으며, 국가안보를 구성하는 여러 영역을 포괄하는 최상위 법이 없는 현재 상태에서 사이버안보를 달성하기 위한 조항만을 독립된 법률로 규율하기 위해서는 입법을 위한 소모적인 노력이 소요될 것으로 예상된다.

반면, 테러방지법에는 위원회, 경보발령, 정보수집, 대책수립 등 기존의 사이버테러 관련 법안에서 규정하고자 했던 사항을 상당수 규율하고 있어, 테러방지법을 개정하여 사이버테러 영역까지 포괄하는 것이 효율적일 것이다. 테러방지법 시행령까지 확대하여 살펴보면 책임기관, 대책본부 운영, 정보공유, 신고 및 포상 등 사이버테러 관련 법안에서 의도한 더 많은 범위를 규정하고 있다. 물론, 현재 테러방지법은 전통적 테러 대응에 중점을 두고 있어 사이버테러 대응과 공통적인 조항은 법률로 규정하고 고유 영역을 하위 규정으로 위임하는 등의 정비도 이루어져야 할 것이다. 그렇게 한다면, 현재 법률적 기반이 부족한 상태에서 대통령령 및 훈령 수준의 사이버테러 (안보) 관련 규정에 대한 기반을 확보할 수 있을 것이다.

이를 위해 가장 먼저 '테러방지법 제2조(정의)'에 사이버테러를 테러 행위의 유형에 포함시켜야 한다. 이때, 앞서 문제점으로 지적한 바와 같이, 사이버테러는 물리적 피해 및 위험을 요구하는 전통적인 테러와는 달리, 보이지 않는 사이버공간에 대한 기밀성, 무결성, 가용성의 훼손이 막대한 파급효과를 미칠 수 있다는 점이 반드시 고려되어야 한다. 반면, 일각에서는 기술 발전에 따라 개인과 사회에 대한 사이버 위협이 점증하는 것이 추세임에도 불구하고, 사이버 보안 문제를 지나치게 거시적인 안보적 시각으로 바라보려 한다거나[18], 사이버테러의 개념을 지나치게 포괄적으로 정의하고 있다고 문제를 제기한다[19]. 따라서, 단순히 침해 (피해)의 규모로

---

18) 고경민, 정영애, 사이버 안보화 문제와 사이버 위협의 포괄적 대응 방안, 한국정보처리학회 학술대회논문집, 24권 1호, 2017, p.363.
19) (재)더미래연구소, 사이버테러방지법의 주요 쟁점분석을 통한 입법 타당성 평가, 국회정보위원회 정책연구개발용역 성과물, 2016,

판단하기보다 테러방지법에 명시된 '테러'의 정의와 국가 안보에 미치는 영향력을 충분히 고려하고, 폭넓은 공감대를 형성한 상태에서 정의가 되어야 한다.

테러방지법을 개정할 경우 사이버테러에 대한 대응도 고전적 형태의 테러에 대한 마찬가지로 피해 확산 방지 및 복구뿐 아니라 테러 행위자 (자산) 에 대한 제압 등 능동적 대응도 허용되어야 할 것이다. 사이버공간에서의 테러는 물리적으로 이격된 장소에서도 이루어질 수 있으므로, 피해 확산 방지 및 복구만으로는 테러 행위자를 제압했다고 할 수 없을 것이다. 따라서, 테러 행위가 이행된 원점에 대한 대응 (역정보 수집, 무력화 등)을 법적으로 허용해야 하겠다.

또한, 조직의 소속 및 임무 수행 범위에 구애받지 않고 보유한 능력을 중심으로 사이버테러 대응 전담조직을 지정하여 대응태세를 유지해야하겠다. 현재 테러방지법과 동법 시행령의 경우 테러 예방 및 대응을 위한 전담조직을 둘 수 있고, 테러가 발생할 경우 현장지휘본부에서 전담조직을 포함한 관계기관의 조직을 지휘 및 통제 할 수 있다고 명시하고 있다. 사이버테러도 조직의 임무영역 및 이해관계에서 벗어나 적시성을 요구하는 사이버테러 대응을 위해 협력하는 분위기를 확산시켜야 할 것이다.

## Ⅴ. 결 론

이미 광범위하게 진행되고 있는 사이버위협의 확산 경향은 전통적인 테러 수행 형태에 변화를 가져올 것이다. 우리나라는 과거부터 사이버테러를 포함한 광범위한 사이버 위협에 대응하기 위한 능력을 확보하고 대응체계를 구축하기 위한 기본법 제정 등 법률적 근거를 확보하고자 오랜 기간 노력해 왔으나 여러 난관에 부딪쳐 결실을 맺지 못하고 있다. 이러한 상황에서 무조건 새로운 법률을 입법 추진하는 것 보다 기존의 테러방지법 및 시행령 개정을 통해 유사한 수준의 기대효과를 달성할 수 있을 것으로 기대한다. 이후 테러 방지 분야를 포함한 포괄적 국가 안보 보장을 위해 필요한 법안에 대한 필요성이 논의되고 성숙된다면 최상위의 안보 관련 기본 법안과 분야별로 파생되는 다양한 법률의 형태로 체계를 구축할 수 있을 것이다.

---

p.18.

09. 사이버테러를 동반한 테러 양상 예측 및 관련 법제도 발전방향# 참 / 고 / 문 / 헌

1. 국내문헌

국가정보원법, 법률 제17646호
국가사이버안전관리규정, 대통령훈령 제316호
국민보호와 공공안전을 위한 테러방지법, 법률 제18321호
사이버안보 업무규정, 대통령령 제31356호
정보통신기반 보호법, 법률 제17357호
정보통신망 이용촉진 및 정보보호 등에 관한 법률, 법률 제 17358호
통합방위법, 법률 제14839호
한국인터넷진흥원, 2021년 사이버 위협 전망, 2021
경찰청 국가수사본부 사이버수사국, "2020 사이버범죄 동향 분석 보고서", 2021
고경민, 정영애, 사이버 안보화 문제와 사이버 위협의 포괄적 대응 방안, 한국정보처리학회 학술대회논문집, 24권 1호, 2017
김성현 및 이창무, "국내 관련 법과 비교 분석을 통한 국가사이버안보법안의 제정 필요성 연구", 한국경호경비학회, 제54호, 2018.
민경식, 김영직, 박진상, 장한나, 랜섬웨어 최신동향 분석 및 시사점, KISA Insight, 한국인터넷진흥원, vol. 2, 2021.
박대로, "여야, 국정원에 광범위 사이버정보 수집권 부여 입법", 뉴시스 (2021.11.19.)
육군사관학교 교수 등 공저, 사이버전 개론, 양서각, 2013
이용석 및 임종인, "사이버 대응태세 구축을 위한 법·제도적 개선방안 연구", 융합보안논문지, 제19권 제1호, 2019.
조상근, "다영역작전 (Multi-Domain Operations, MDO) - 지상, 해상, 공중, 우주, 사이버·전자기 영역을 종횡무진(縱橫無盡)하는 미 육군의 미래 작전수행개념", https://bemil.chosun.com/site/data/html_dir/2020/12/31/2020123100581.html (검색일: 2021.10.24.)
(재)더미래연구소, 사이버테러방지법의 주요 쟁점분석을 통한 입법 타당성 평가, 국회정보위원회 정책연구개발용역 성과물, 2016KOREAN NATIONAL POLICE AGENCY    227

2. 외국문헌

A / B / S / T / R / A / C / T

# Prediction of terrorism pattern accompanied by cyber-terrorism and the development direction of corresponding legal systems

As the information and communication system has become an essential element for national operation and people's lives, and the dependence on information and communication systems such as national infrastructure systems and facilities increases, cyber terrorism is rapidly emerging as a serious threat to national security in peacetime. As terrorist groups' access to cyber-attack assets improves, the traditional form of terrorism is also expected to change to a form combined with cyber-terrorism. Nevertheless, from a national security point of view, Korea lacks a legal system to prepare for and respond to cyber terrorism. In this paper, based on the development process of the modern military operation concept, we predict the changes in the form of terrorism, analyse the restrictions on the national response to cyber-terrorism based on the current legal system, and propose the development directions.

➲ **Key words**
cyber-terrorism, future terrorism patterns, national security, legal systems